\documentclass[twocolumn]{aastex631}
\usepackage{color}
\usepackage{amsmath}

\newcommand{\ie}{i.e.,\ }
\newcommand{\eg}{e.g.,\ }
\newcommand{\etal}{et~al.\ }

\newcommand{\magsec}{mag arcsec$^{-2}$}
\newcommand{\Msun}{$M_\odot$}

\newcommand{\ha}{H$\alpha$}

\begin{document}

\title{BST1047+1156: A (Failing) Ultradiffuse Tidal Dwarf in the Leo I Group}
\shorttitle{Stellar Populations in BST1047+1156}
\shortauthors{Mihos \etal}

\author{J. Christopher Mihos}
\affiliation{Department of Astronomy, Case Western Reserve University, Cleveland OH 44106, USA}

\author{Patrick R. Durrell}
\affiliation{Department of Physics, Astronomy, Geology, and Environmental Sciences, Youngstown State University, Youngstown, OH 44555 USA}

\author{Aaron E. Watkins}
\affiliation{Centre for Astrophysics Research, University of Hertfordshire, College Lane, Hatfield AL10 9AB, UK}

\author{Stacy S. McGaugh}
\affiliation{Department of Astronomy, Case Western Reserve University, Cleveland OH 44106, USA}

\author{John Feldmeier}
\affiliation{Department of Physics, Astronomy, Geology, and Environmental Sciences, Youngstown State University, Youngstown, OH 44555 USA}

\begin{abstract}

We use deep {\sl Hubble Space Telescope} imaging to study the resolved
stellar populations in BST1047+1156, a gas-rich, ultradiffuse dwarf
galaxy found in the intragroup environment of the Leo I galaxy group.
While our imaging reaches approximately two magnitudes below the tip of
the red giant branch at the Leo I distance of 11 Mpc, we find no
evidence for an old red giant sequence that would signal an extended
star formation history for the object. Instead, we clearly detect the
red and blue helium burning sequences of its stellar populations, as
well as the fainter blue main sequence, all indicative of a recent burst
of star formation having taken place over the past 50--250 Myr.
Comparing to isochrones for young metal-poor stellar populations, we
infer this post-starburst population to be moderately metal poor, with
metallicity [M/H] in the range $-1$ to $-1.5$. The combination of a
young, moderately metal-poor post starburst population and no old stars
motivates a scenario in which BST1047 was recently formed during a weak
burst of star formation in gas that was tidally stripped from the
outskirts of the neighboring massive spiral M96. BST1047's extremely
diffuse nature, lack of ongoing star formation, and disturbed HI
morphology all argue that it is a transitory object, a ``failing tidal
dwarf'' in the process of being disrupted by interactions within the Leo
I group. Finally, in the environment surrounding BST1047, our imaging
also reveals the old, metal-poor ([M/H]$=-1.3\pm0.2$) stellar halo of
M96 at a projected radius of 50 kpc.

\end{abstract}


\section{Introduction}

The properties of extreme low surface brightness (LSB) galaxies continue
to challenge models of galaxy formation and evolution. While much
attention has been focused recently on the ``ultra-diffuse galaxies''
found in dense galaxy clusters, gas-rich LSBs found in the field and
group environments \citep[\eg][]{mcgaugh94, cannon15, leisman17} may
have a less complicated evolutionary path, and better probe mechanisms
driving galaxy formation at the lowest densities. For example, the high
gas fractions and low metallicities of LSB galaxies \citep{mcgaugh94,
ellison08, pilyugin14} argue that they have converted little of their
baryonic mass into stars. This is likely due to their extremely low gas
densities \citep{vanderhulst93, vanzee97, wyder09}, which result in a
sputtering and inefficient star formation history \citep{schombert01,
schombert14, schombert15}. Thus, these galaxies raise questions both
macro and micro: how galaxy formation is linked to the global
environment, and how stars form on smaller scales within galaxies.

The recent discovery of the extreme LSB galaxy BST1047+1156
\citep[hereafter BST1047]{mihos18bst} is particularly notable in this
context. With an HI velocity that places it unambiguously within the
Leo~I galaxy group \citep[D=11 Mpc;][]{graham97, lee16}, BST1047 has the
{\sl lowest} surface brightness of any known star forming galaxy
($\mu_{B,peak}$=28.8 \magsec), an isophotal radius of $R_{30} \approx$ 2
kpc, and a total gas mass of $4.5\times10^7$ \Msun \citep{mihos18bst}.
The object's {\sl peak} HI column density ($1.4\times 10^{20}\ {\rm
cm}^{-2}$) is well below that in which stars typically form
\citep{bigiel08, bigiel10, krumholz09, clark14}, yet its extremely blue
optical colors ($B-V=0.14 \pm 0.09$) and GALEX far-UV emission both
argue for the presence of young stars \citep{mihos18bst}. BST1047's
combination of extraordinarily high gas fraction ($f_g \approx0.99$),
extremely blue optical colors, and vanishingly low surface brightness
makes it the most extreme gas-rich LSB object known to date.

Exactly how BST1047 formed, and what has triggered its recent star
formation, remains unclear. The Leo~I group is awash in extended HI,
including the large ``Leo HI Ring'' surrounding NGC~3379 to the north
\citep{schneider85, schneider86}, likely a remnant of past tidal
interactions \citep{micheldansac10, corbelli21}. BST1047 itself is
embedded in a low density HI stream connecting the Ring to the spiral
galaxy M96. This, plus the fact that BST1047 sports a pair of HI tidal
tails of its own, suggests the object may be an extremely diffuse LSB
galaxy recovering from a weak burst of tidally-triggered star formation.
Alternatively, BST1047 may be a ``tidal dwarf galaxy,''
\citep[\eg][]{duc00, lelli15}, spawned {\sl directly} from tidally
compressed gas, with the young stars marking its formation age. Since
tidal dwarfs should be free of dark matter \citep{barnes92, elmegreen93}
and perhaps only tenuously bound, under this scenario BST1047 may be a
short-lived object~--- a ``failing'' tidal dwarf caught in the throes of
tidal disruption in the group environment.

\begin{figure*}[]
\centerline{\includegraphics[width=7.0truein]{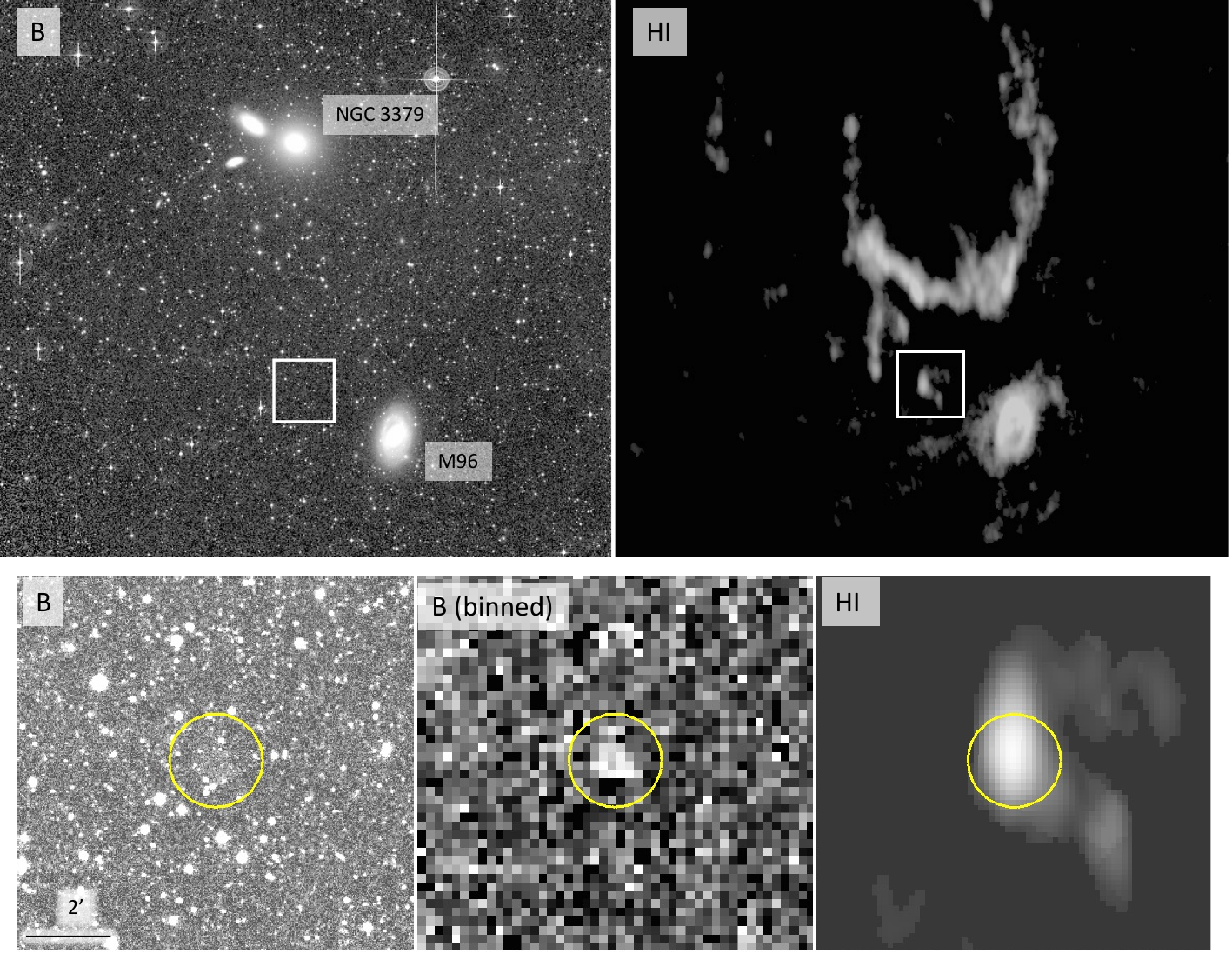}}
\caption{Optical and HI imaging of the Leo I Group and BST1047+1156,
taken from \citet{mihos18bst}. The upper left panel shows the deep
wide-field B-band imaging of \citet{watkins14}, while the upper right
panel shows the HI map from \citet{oosterloo10}. In these panels, the
white box shows the location of BST1047 in the Leo I group, and is blown
up in the lower panels. The lower left panel shows the B-band image,
while the lower center panel shows the B-image after being masked of
compact sources and rebinned in 9$\times$9 pixel boxes to show low
surface brightness emission. The lower right panel shows the HI map on
the same scale. In the lower panels, the yellow circle is 70\arcsec\ in
radius, twice the size of the $R_{30}$ isophote. }
\label{optHI}
\end{figure*}

Either of these scenarios has important ramifications for issues
surrounding theories of formation and evolution of low mass galaxies. If
BST1047 is a diffuse but long-lived LSB galaxy, with an established, old
stellar population, it challenges star formation models which posit that
stars should not form at such low gas densities. Under such models,
where has the older population come from? How can galaxies this diffuse
sustain such prolonged star formation histories? Conversely, if BST1047
is a disrupting tidal dwarf, it would provide insight into the
evolutionary link between tidal interactions, formation and disruption
of dwarf galaxies, and the deposition of young stars into the intragroup
medium. Key to resolving the question of BST1047's origin is an
understanding of its stellar populations --- in particular, does it have
a well established old red giant branch sequence, indicative of an
long-lived star forming history, or are the stellar populations
exclusively young, as might be expected BST1047 was recently formed
during a tidal encounter?

To answer these questions, we use deep {\sl Hubble Space Telescope} ACS
imaging to study the stellar populations of BST1047. Using the F606W and
F814W filters, our imaging extends roughly two magnitudes below the
expected tip of the red giant branch at the Leo I distance, allowing us
to detect and characterize stellar populations across a range of ages,
including any red giant branch stars, red and blue helium burning stars,
and potentially even upper main sequence stars. These various
populations give constraints on the ages and the metallicities of both
young and old stellar populations, providing strong constraints on the
extended star formation history in BST1047.

\section{Observational Data}

\subsection{Imaging and Reduction}

\begin{figure*}[]
\centerline{\includegraphics[width=7.0truein]{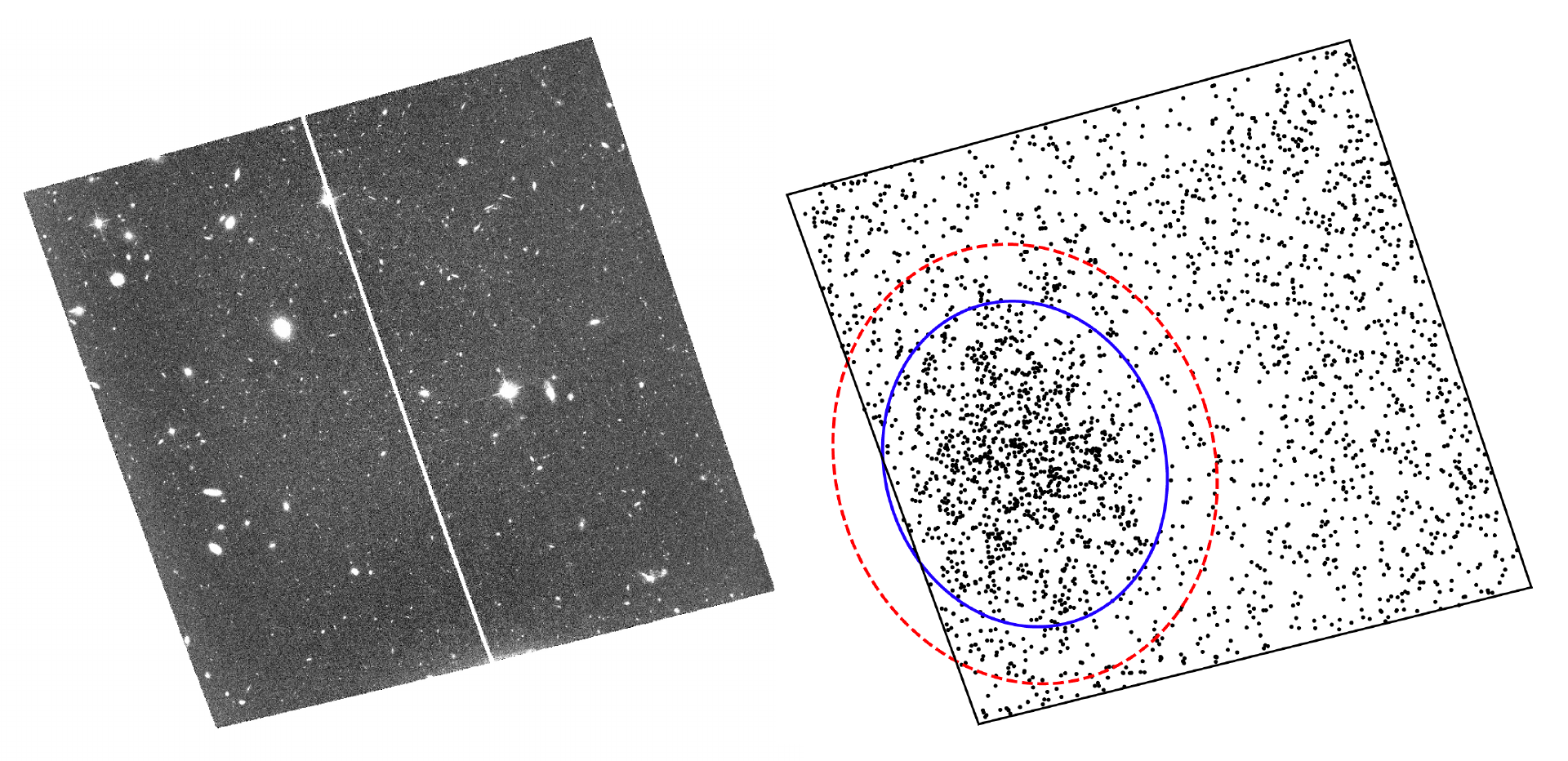}}
\caption{Left panel: Stacked F814W ACS image of BST1047+1156, with a 
total exposure time of 14$\times$1215s. North is up, east is to the left,
and the field of view is 202\arcsec$\times$202\arcsec. Right panel: Spatial
distribution of point sources detected in the ACS imaging. All point sources
within the solid blue ellipse are used for the analysis of BST1047, while point sources
found outside the dashed red ellipse are considered part of the background or
surrounding environment.
}
\label{ACSimage}
\end{figure*}

We imaged BST1047 using the Wide Field Channel (WFC) of the Advanced
Camera for Surveys (ACS) on the Hubble Space Telescope (HST) under
program GO-16762. The imaging field, shown in the left panel of
Figure~\ref{ACSimage}, places BST1047 in the eastern side of the ACS
field of view, avoiding nearby bright stars and leaving the western side
blank for background estimation.

The field was imaged over 8 orbits in F606W and 7 orbits in F814W; each
orbit consisted of two 1185s exposures, yielding total exposure times of
16458s and 16590s in F606W and F814W, respectively (one F606W exposure
was cut short due to guide star loss). Each visit made use of a small
($\sim$ 3.5--4.5 pixel) custom four point box dither pattern to aid in
sub-pixel sampling of the ACS images and to also avoid placing any
objects on bad or hot pixels. The different visits were further shifted
in slightly larger (20 pixel) offsets to avoid other artifacts and
facilitate effective cosmic ray removal in our long (1/2 orbit)
exposures. As BST1047 is small enough to fit into a single WFC chip, the
galaxy was centered on the WFC1 chip, and no attempt was made to cover
the ACS chip gap.

Point-source photometry is carried out on the individual, CTE-corrected
{\tt flc} images using DOLPHOT (described below), which requires a
sufficiently deep drizzled image to use as an astrometric reference. To
create this image, the individual images from different visits needed to
be precisely aligned. We found that images from the three visits
(including eight F606W images and two F814W images) that were
astrometrically calibrated with the GSC v2.4.2 catalog were slightly
offset ($\sim0.5$pixel in F606W; $\sim 0.2$pixels in F814W) from the
remaining 20 images calibrated to the newer Gaia eDR3 catalog. To
improve the relative image alignments, we used the {\tt
drizzlepac/tweakreg} package to adjust the image world coordinate
systems based on point source positions on the individual {\tt flc} images
measured using Source-Extractor \citep{source-ext}. After these
corrections, we used {\tt drizzlepac/astrodrizzle} to create stacked
deep F606W and F814W images of the ACS field; the F814W image is shown
in the left panel of Figure~\ref{ACSimage}.

\subsection{Point Source Photometry and Artificial Star Tests}

With such an extremely low surface brightness \citep[$\langle \mu_B
\rangle_e=28.9$ \magsec; ][]{mihos18bst}, the integrated light from
BST1047 is too faint to show up unambiguously in our ACS imaging;
instead, we only detect it through its resolved stellar populations. We
use the software package DOLPHOT \citep[an updated version of
HSTPhot;][]{dolphin00} to perform point-source photometry of objects on
the individual CTE-corrected {\tt flc} images using pre-computed Tiny
Tim PSFs \citep{krist95}. We performed object detection and photometry
on all 30 individual images (16 in F606W; 14 in F814W) images
simultaneously, using the deep F814W drizzled image stack created above
for the reference image. We used the Nov 2019 version of DOLPHOT~2.0
(available at http://americano.dolphinsim.com/dolphot/) to pre-process
the raw {\tt flc} images, applying bad-pixel masks and pixel-area masks
({\tt acsmask}), splitting the images into the individual WFC1/2 chip
images ({\tt splitgroups}), and constructing an initial background sky
map for each chip/image from each image ({\tt calcsky}).

Photometry with DOLPHOT is very dependent on the choice of input
parameters \citep[see ][]{williams14}, so we experimented with a number
of the parameters, finally settling on values similar to those used in
previous deep photometric studies with ACS
\citep[\eg][]{williams14,mihos18m101,shen21} and/or suggested by the
DOLPHOT/ACS User's Guide. As our ACS field is relatively uncrowded, we
adopted a photometric aperture {\tt RAper}=4.0 pix, a PSF fitting region
of {\tt RPSF}=10 pix, and the {\tt FITSKY}=1 option for derivation of
the sky background The only changes we made to the usual DOLPHOT
workflow were the derivation of the aperture corrections on each
chip/image. With so few bright stellar objects in our frames, some of
the individual DOLPHOT-computed aperture corrections (and thus the
individual F606W/F814W magnitudes) could be affected, even for brighter
stars. To improve this, we input our own visually-selected list of 53
isolated stellar objects over the entire field which DOLPHOT could use
to compute aperture corrections. The final aperture corrections for each
chip/image/filter were based on anywhere from 6 to 29 measured stars.
Finally, the instrumental magnitudes were converted to the VEGAMAG {\sl
HST} photometric system. We used updated zeropoints (at the time of
observations, using the ACS zeropoint calculator
(https://acszeropoints.stsci.edu/) of 26.398 for F606W and 25.502 for
F814W. We present all photometry in the VEGAMAG system unless explicitly
stated otherwise.

To ensure the most accurate point source photometry, we apply the
following selection parameters to the photometric catalog. We start by
selecting only those objects with DOLPHOT object TYPE=1 (``good star'')
and signal-to-noise S/N$>$3.5 in {\it both} the F606W and F814W filters.
We also only select sources that are uncrowded (CROWD$<$0.25) and have a
goodness of fit value of CHI$<$2.4 in both filters; these values are
based both on visual inspection of bright stars and galaxies in our
images, and the results from the artificial stars detailed below. At
fainter magnitudes (F814W$>$26), contamination from unresolved
background galaxies becomes problematic. To reduce this contamination,
we also make a magnitude-dependent cut on the DOLPHOT SHARP parameter,
using $| SHARP | < 0.04 + 0.3e^{(m-m_{\rm crit})}$, with $m_{\rm crit}
=$ 29.5 and 28.7 in F606W and F814W, respectively. This function is
similar to that used in the our previous {\it HST} studies of stellar
populations in M101 and the Virgo Cluster \citep{mihos18m101, mihos22},
and the function parameters are chosen based both on the observed
photometric catalog and on our artificial star analysis. We have also
checked that the sources rejected under our SHARP criteria do not show
stellar population-like patterns in the color-magnitude diagram that
might suggest we are over-aggressively rejecting actual stars in the Leo
I group environment. The spatial distribution of point sources selected
in this fashion are shown in the right panel of Figure~\ref{ACSimage}.

\begin{figure*}[]
\centerline{\includegraphics[width=7.0truein]{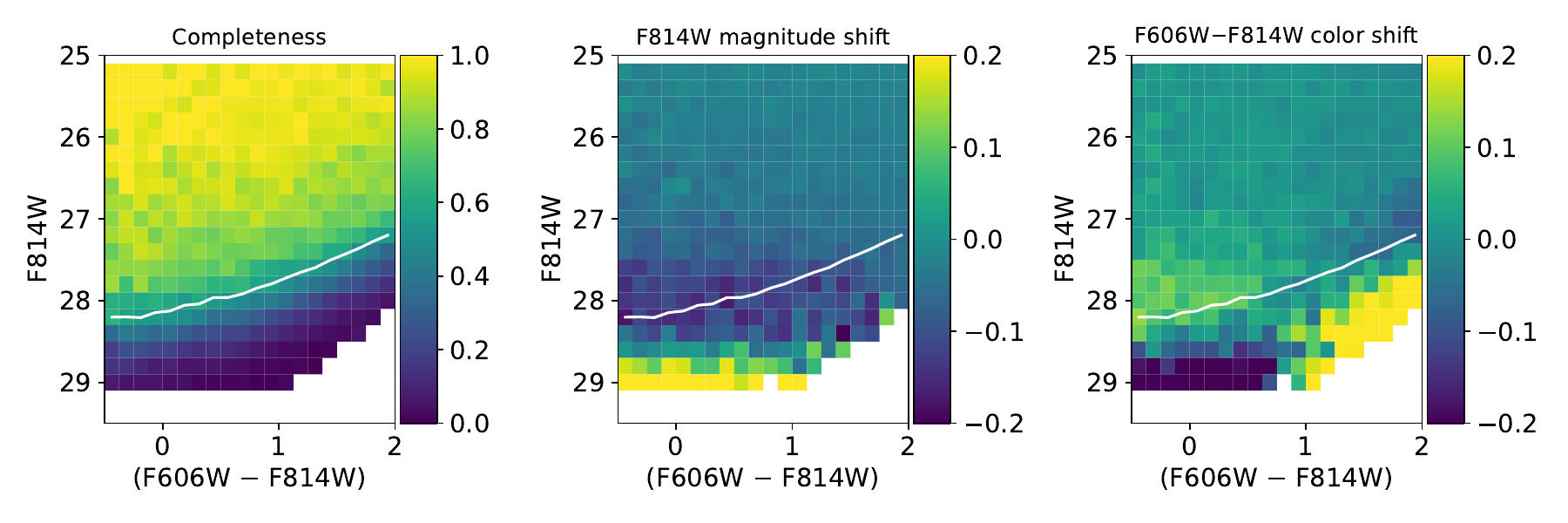}}
\caption{Results of artificial star tests. Left panel: Completeness,
Middle panel: F814W magnitude shift, Right panel: F606W$-$F814W
color shift. Shifts are measured as input minus measured values,
such that a positive magnitude shift corresponds to a star being
measured systematically too bright, and a positive color shift
corresponds to a star being measured systematically too blue.
In each panel, the white line shows the 50\% completeness limit.
}
\label{artstars}
\end{figure*}

To assess the photometric completeness and bias of the photometry in our
ACS imaging, we use DOLPHOT to insert and measure 100,000 artificial
stars over the magnitude range $22 < {\rm F606W} < 30$ and color range
$-0.5 < {\rm F606W-F814W} < 2.0$. We process the artificial stars using
the same photometric selection criteria used for the actual data, and
plot in Figure~\ref{artstars} the completeness fraction and shift in
magnitude and color (defined as input minus measured) as a function of
F814W magnitude and F606W$-$F814W color. Because of our joint selection
in F606W and F814W, completeness is a function of both magnitude and
color, with 50\% completeness at F814W=28.2 in the blue (at
F606W$-$F814W=0.0), and rising to F814W=27.8 in the red (at
F606W$-$F814W=1.0). At magnitudes brighter than F814W=27.0 we see little
systematic shift in either magnitude or color, but at fainter magnitudes
shifts in both are evident at the $\approx$ 0.1 mag level, consistent
with our previous analysis of ACS data in \citet{mihos18m101}. In our
analysis that follows, we always plot magnitudes and colors as measured,
correcting only for foreground extinction \citep[$A_{\rm F606W}=0.062,
A_{\rm F814W}=0.038$,][]{schlafly11}, and use the results of the
artificial star tests to adjust the theoretical stellar isochrones to
account for these systematic effects when interpreting our photometric
results.

\section{Analysis}

\begin{figure*}[]
\centerline{\includegraphics[width=7.0truein]{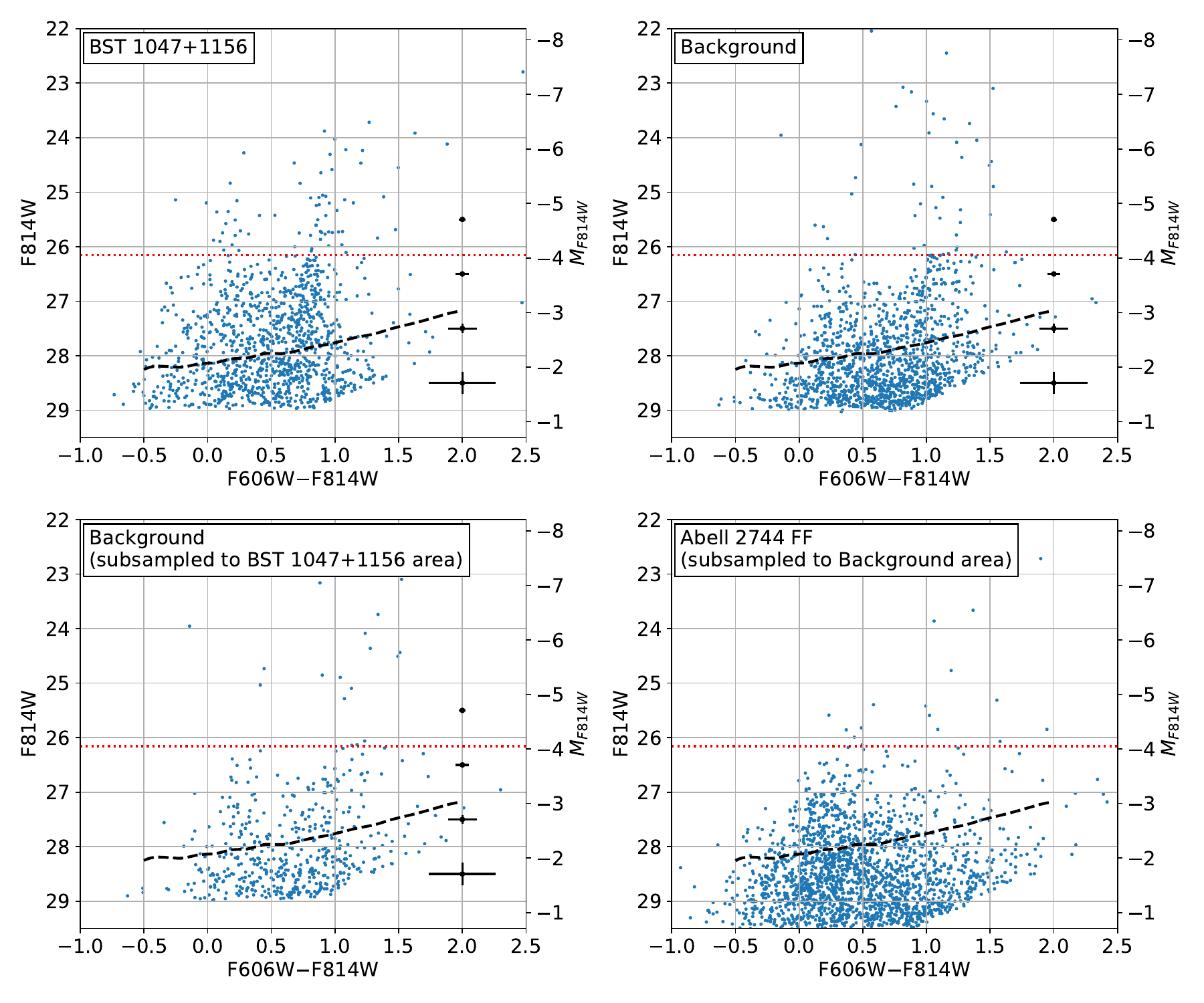}}
\caption{Color-magnitude diagrams for point sources in the ACS imaging.
The top panels shows the extracted CMDs for BST1047 (left) and the
background (right). The lower left panel shows the background CMD
randomly subsampled to match the area of BST1047, while the lower right
panel shows the point source CMD for the Abell 2744 Flanking Field data
(taken from the analysis of \citealt{mihos18m101}), randomly subsampled to
match the area of the background field. Each CMD in the lower panels
thus acts as a control field for the CMD immediately above it. In each
subpanel, the right axis shows the apparent magnitude, the left axis
shows the absolute magnitude at the adopted 11.0 Mpc distance of the Leo
I Group, and the red dotted line shows the expected RGB tip magnitude.
Typical errorbars as a function of magnitude are shown in black, and the
50\% completeness limit for the BST1047 imaging is shown as the dashed
black line. }
\label{rawcmds}
\end{figure*}

The right panel of Figure~\ref{ACSimage} shows the spatial distribution
of point sources in our ACS field; an excess of sources corresponding to
the stellar population of BST1047 can clearly be seen on the eastern
half of the FOV. The distribution of point sources appears slightly
elongated roughly along the north-south axis, and shows small scale
clumpiness as well. We construct a color-magnitude diagram (CMD) for
BST1047 by extracting all point sources within an ellipse (determined by
eye and shown in Figure~\ref{ACSimage}) centered at
$(\alpha,\delta)_{J2000}$=(10:47:43.59, 11:55:47.0), and having an
ellipticity of 0.85, semimajor axis of 50\arcsec, and position angle of
17\degr. The center of this ellipse is approximately 14.3\arcsec\ south
of BST1047's center coordinate originally reported in
\citet{mihos18bst}. For comparison, we also construct a background CMD
by extracting sources that lie outside a 350 pixel (17.5\arcsec) buffer
around the BST1047 ellipse, shown as the dotted red ellipse in
Figure~\ref{ACSimage}. The extracted CMDs for each region (BST1047 and
background) are shown in the top panels of Figure~\ref{rawcmds}.

While the background region is meant as a control for the BST1047 field,
it has a much larger area (by a factor of 2.87), and thus
over-represents the potential contamination to BST1047's CMD. The lower
right panel corrects for this difference in area by randomly subsampling
sources in the background region by a factor of 2.87 to match the area
of the BST1047 field, thus acting as a more representative control
sample for BST1047.

Of course, the background region itself is not a pure background. As
BST1047 resides within the Leo I group, and also sits projected only
15\arcmin\ (48 kpc) northeast of the luminous spiral galaxy M96, our ACS
pointing samples not only background sources, but also stars in M96's
extended stellar halo, as well as any potential Leo~I intragroup stars
\citep{watkins14, ragusa22}. To estimate a cleaner background CMD, we
turn to the deep HST imaging of the Abell 2744 Flanking Field
\citep{lotz17}. That imaging used the same filters used here, and in
\citet{mihos18m101} we extracted point source photometry for that
imaging using the same techniques as described above. Thus, it acts as a
reasonable control field for our background region here. In the lower
right panel of Figure~\ref{rawcmds} we show the Abell 2744 Flanking
Field photometry, using the same selection criteria as used in this
study, and subsampled down by a factor of 1.62 to match the area of our
background region.

We start with a discussion of the CMD in the full background region
(upper right panel of Figure~\ref{rawcmds}), comparing it to its control
field, the subsampled Abell 2744 FF field directly below it. The most
striking feature of the background region is the clear signature of a
metal-poor red giant branch population, terminating at the expected RGB
tip at F814W=26.2. Brighter than this, there are a number of red stars
in the field, possibly AGB stars or true background contaminants. At
these magnitudes, and over the small ACS field of view, foreground
contamination from Milky Way stars should be small; comparing to the
TRILEGAL models of \citet{girardi05, girardi16} we would expect only a
handful of objects brighter than the observed RGB tip. At fainter
magnitudes (F814W$>$26) we also see a swarm of bluer sources with colors
$0.0 < {\rm F606W-F814W} < 0.5$, but these sources appear comparable in
number to those seen in the Abell 2744 Flanking Field, and are likely
unresolved background sources. Finally, we also see a handful of
brighter sources in this bluer color range, but not obviously in excess
of the background expectation.

Turning to the CMD for BST1047 itself, we again see a clear red sequence
of stars, but one that is distinctly bluer than that in the background
region. Whereas the sequence in the background region reaches a color of
F606W$-$F814W$\approx$1.03 when it reaches the RGB tip, the sequence in
BST1047 has a color of F606W$-$F814W$\approx$0.78 at a comparable
brightness, and continues on to brighter magnitudes above the
expectation for the RGB tip.

\begin{figure}[]
\centerline{\includegraphics[width=3.5truein]{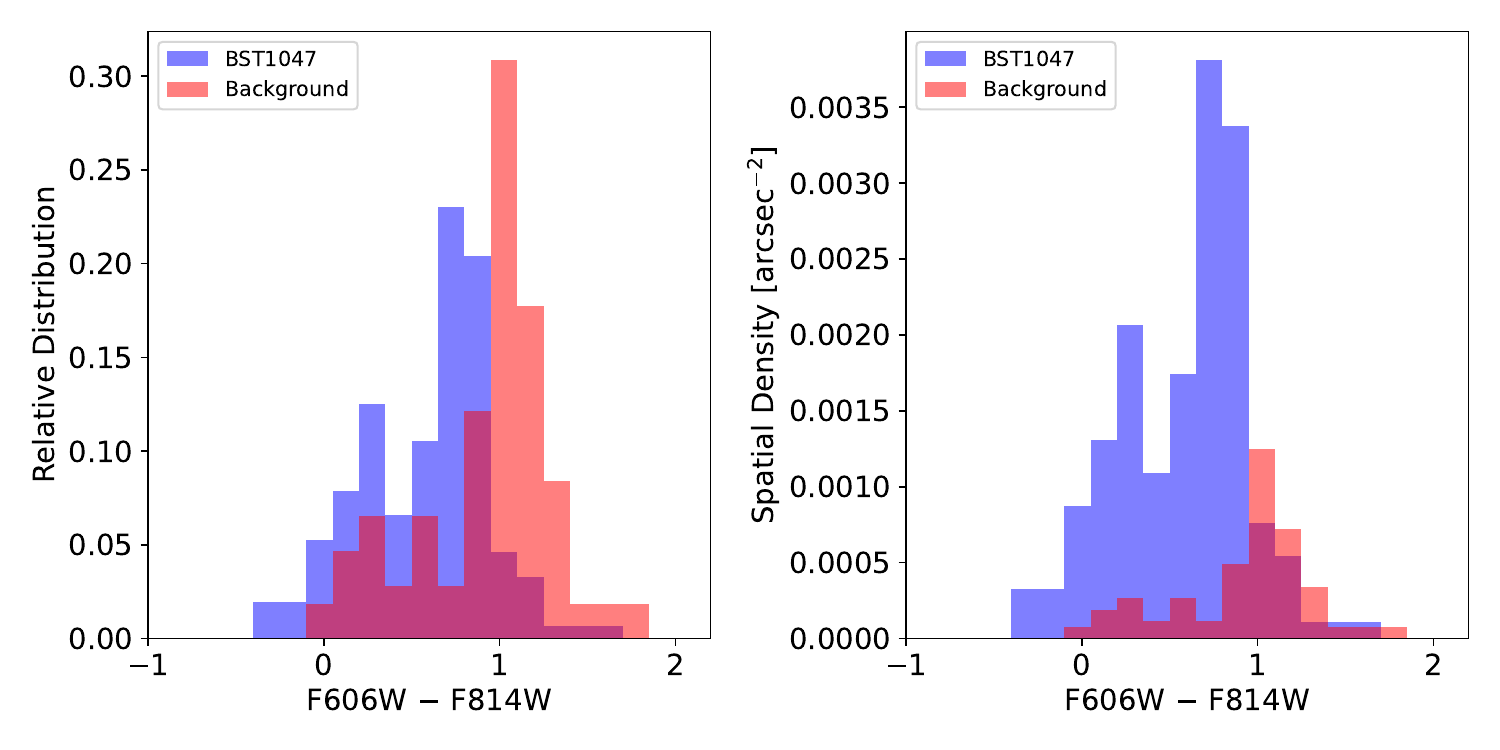}}
\caption{Color distribution of point sources in the magnitude range
$m_{\rm tip,F814W} < {\rm F814W} < m_{\rm tip,F814W}+0.75$. The left
panel shows the relative fraction of sources in each field, while the
right panel shows the actual number density of sources.
}
\label{colordist}
\end{figure}

We demonstrate this color difference in Figure~\ref{colordist}, which
shows the color distribution of point sources of all colors, in the
magnitude range $m_{\rm tip} \leq {\rm F814W} \leq m_{\rm tip}+0.75$
(\ie within 0.75 magnitudes of the expected RGB tip). The left panel
shows the relative color distribution in each region, where the color
difference between the red sequences in each region is clear. The right
panel of Figure~\ref{colordist} shows the surface density of sources as
a function of color -- in other words, the color distribution normalized
by area. Here too, the difference in the red sequences is dramatic: not
only are they different in color, the density of red stars is much
higher in BST1047 than in the background. The two sequences are clearly
tracing different populations of stars.

At magnitudes brighter than the RGB tip, the CMDs in
Figure~\ref{rawcmds} show an excess of stars in BST1047 both at red
colors (F606W$-$F814W $\approx$ 0.8--1.0) and in the blue (F606W$-$F814W
$\approx$ 0.0--0.4) compared to the background region. The morphology
and color of these bright red and blue sequences suggest they are helium
burning sequences from evolving massive stars, signatures of recent star
formation in BST1047. At fainter magnitudes (F814W $\sim$ 26--28) we
also see an excess population of stars with very blue colors of
F606W$-$F814W $<$ 0.0 compared to the background. These sources can also
be seen in the color distributions shown in Figure~\ref{colordist}, and
may represent massive stars still on the upper main sequence.

\begin{figure*}[]
\centerline{\includegraphics[width=7.0truein]{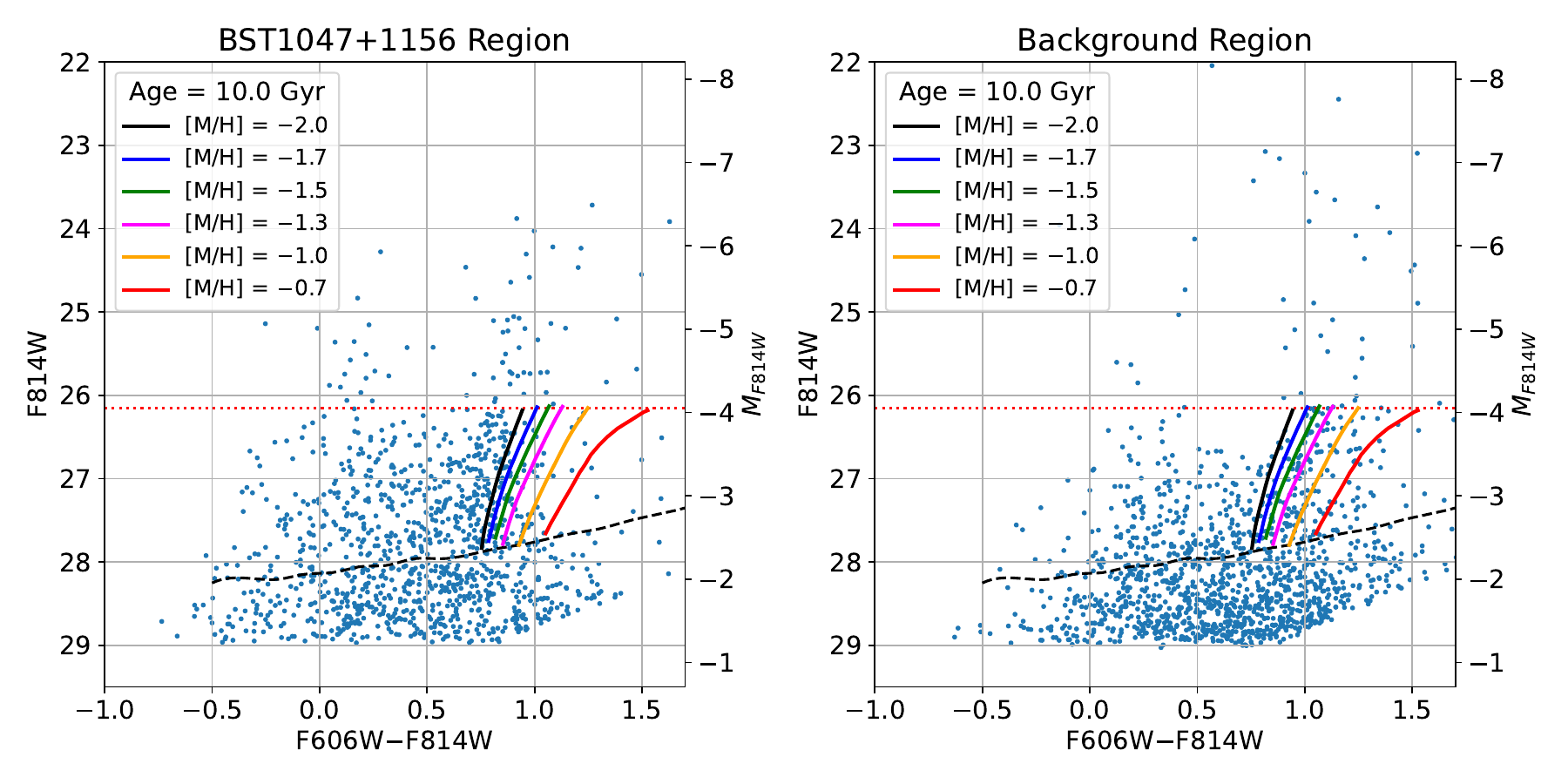}}
\caption{CMDs for BST1047 (left) and background (right) regions,
overlaid with PARSEC 1.2S isochrones for 10 Gyr old populations of
varying metallicities. These isochrones have been adjusted to reflect
the small systematic photometric shifts in magnitude and color in our
data as determined from the artificial star tests (see Section 2.2), but
in the red these shifts are negligible at magnitudes brighter than 
F814W $\approx$ 27.5.
}
\label{oldtracks}
\end{figure*}

In Figure~\ref{oldtracks} we overlay isochrones for old stellar
populations of varying metallicities onto the CMDs for both BST1047 and
the background regions. We use the PARSEC 1.2S isochrones
\citep{bressan12, marigo17}, with a fixed age of 10 Gyr, and with a
range of metallicities spanning [M/H]=$-2$ to $-0.7$. We adjust the
tracks to reflect the systematic photometric shifts discussed in \S2.2,
but note that in this portion of the CMD the shifts are negligible at
the RGB tip, and always $<$0.02 mag even down at the 50\% completeness
limit. Looking at the background region, which likely includes
populations in M96's outer stellar halo, the red sequence seen there is
well-matched by old RGB tracks with metallicity [M/H] in the range $-1$
to $-1.5$. Presuming this is M96's halo we are seeing, and adopting a
typical halo alpha abundance of $[\alpha/{\rm Fe}]=+0.3$, the
metallicity corresponds to [Fe/H]~$ \approx -1.2$ to $-1.7$
\citep{salaris93,streich14}. These metallicities are similar to
those found in the outer halos of nearby spirals in the GHOSTS project
\citep{monachesi16}, again arguing these stars belong to the old halo
population of M96. In contrast, the old isochrones provide a poor match for
the red sequence in BST1047; the most metal-poor isochrone
([M/H]$=-2.0$) only reaches a color of F606W$-$F814W = 0.9,
significantly redder than the mean color of the red sequence in BST1047
(F606W$-$F814W = 0.76). This red sequence in BST1047 likely then
consists of young red helium burning stars or very metal poor
intermediate age RGB stars.

\begin{figure*}[]
\centerline{\includegraphics[width=7.0truein]{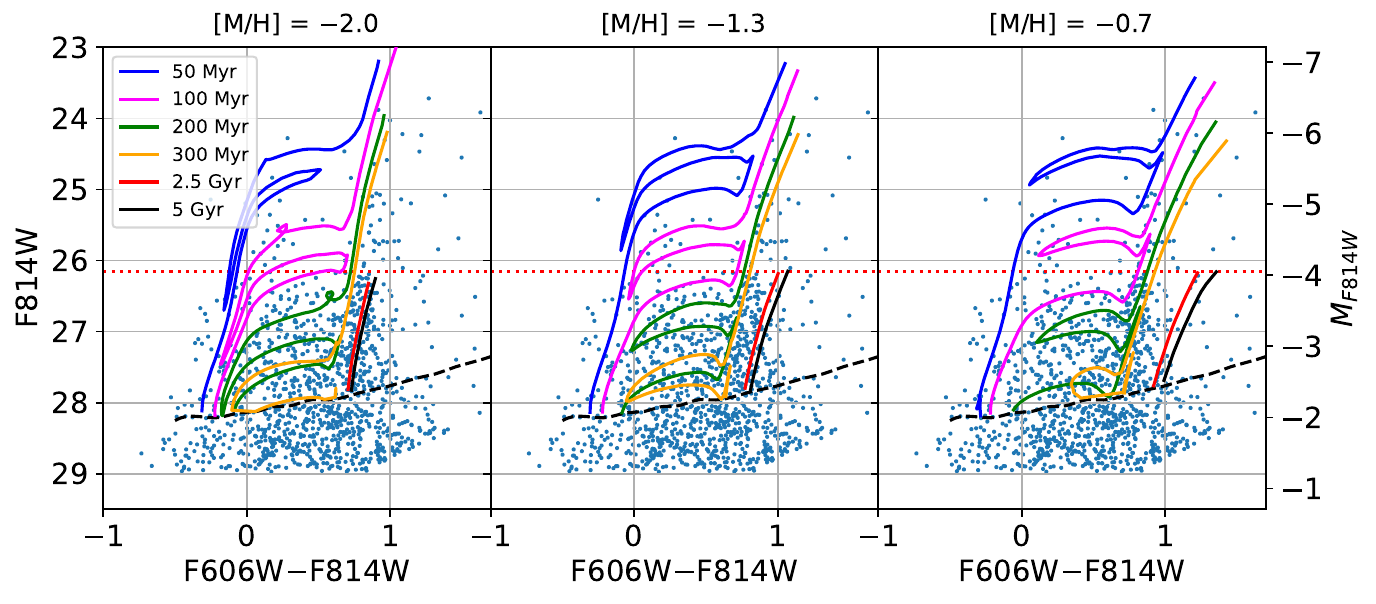}}
\caption{CMDs for BST1047 overlaid with PARSEC isochrones of stellar
populations with varying ages and metallicities of [M/H]=$-2.0$ (left),
$-1.3$ (middle), and $-0.7$ (right). These isochrones have been adjusted
to reflect the small systematic photometric shifts in magnitude and
color in our data as determined from the artificial star tests (see
Section 2.2). These shifts in the isochrones are most noticeable as a
slight blue ``bulge'' in the tracks at F814$>27.5$ and F606W$-$F814W
$<0$.}
\label{youngtracks}
\end{figure*}

We compare the CMD of BST1047 to younger isochrones in
Figure~\ref{youngtracks}, which overplots the PARSEC 1.2S isochrones for
a range of young and intermediate ages, using metallicities of [M/H] =
$-2.0$, $-1.3$, and $-0.7$. We again adjust the isochrones for the
systematic photometric shifts. At bright magnitudes (F814W$<$27) the
shifts remain negligible, but in the blue at fainter magnitudes
(F606W$-$F814W$<$0, F814W$>$27, the systematic blueward shift becomes
more noticeable, shifting the isochrones bluer by $\approx 0.05-0.1$ mag
near the 50\% completeness limit and leading to the slight ``bluish
bulge'' of the tracks in this region. With these effects in mind, these
tracks show that the most luminous stars, 1--2 magnitudes brighter than
the RGB tip, are consistent with blue and red helium burning sequences
arising from massive stars younger than a few hundred million years old.
At fainter magnitudes, the population of objects with very blue
F606W$-$F814W colors $< 0.0$ may be tracing massive main sequence stars
as young as 50 Myr. Looking at the intermediate age RGB tracks, even at
younger ages, RGB sequences are still generally too red to match the red
sequence we see in BST1047, except perhaps at the very lowest
metallicity ([M/H]$=-2$) and with relatively young ($<$2.5 Gyr) RGB
populations. However, given the clear detection of BHeB stars in the
same region, the most natural explanation for the red sequence in BST
1047 is that it is the associated RHeB sequence, with little evidence
for a significant population of old RGB stars in the field.

In terms of metallicity, the sparseness of the stellar populations and
the tight spacing of tracks within the helium burning sequences make it
hard to give tight metallicity constraints for the population.
Nonetheless, the stars are clearly metal poor, with [M/H] $\approx -1.0$
or somewhat lower. More metal-rich than this, the red helium burning
sequences turn much redder than observed in BST1047 where the sequence
remains bluer than F606W$-$F814W = 1.0. At the most metal-poor extreme,
[M/H]=$-2.0$, both the red and blue helium burning tracks start to shift
bluer than seen in the observed CMDs, making it unlikely that the
populations are this metal-poor.

\begin{figure*}[]
\centerline{\includegraphics[width=7.0truein]{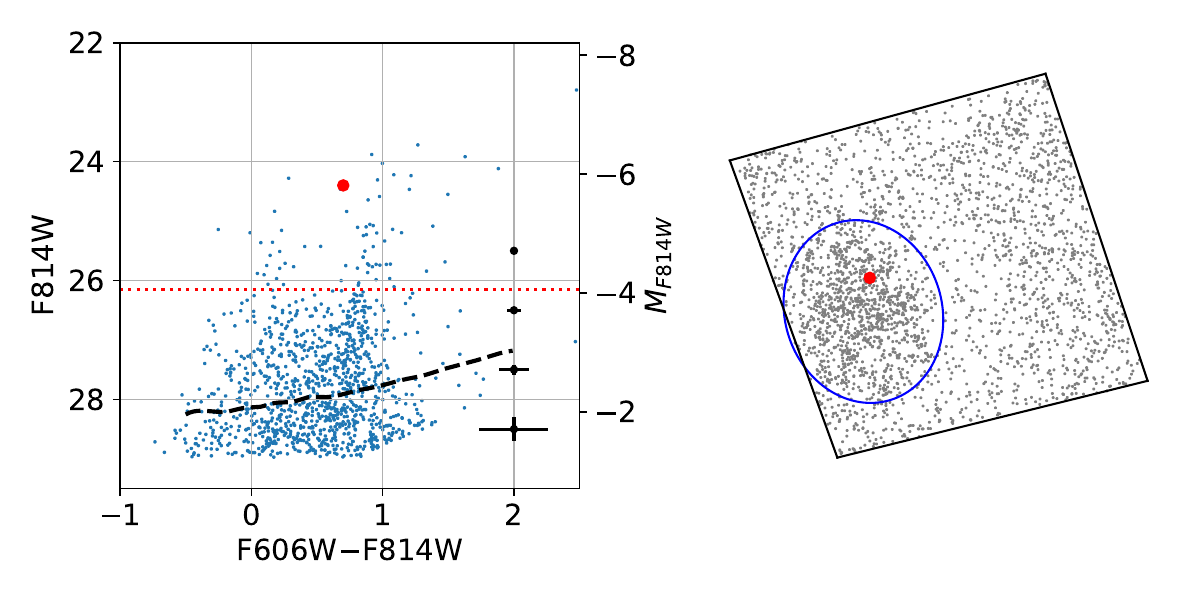}}
\caption{The CMD location (top) and spatial position (bottom) of the
bright variable star detected in our imaging. In both plots, the red
symbol shows the location of the variable. The CMD shows all point
sources within the BST1047 region, which is shown by the blue
oval in the lower panel. Lines and symbols in the CMD are as 
described in Figure~\ref{rawcmds}, and the lower figure shows
the 202\arcsec$\times$202\arcsec ACS field of view, with North
up and East to the left.}
\label{variable}
\end{figure*}

In our photometry, we also find one variable star with properties
potentially consistent with being a luminous Cepheid variable. The
source is located 20\arcsec\ north of the center of BST1047, at
$(\alpha, \delta)_{J2000}$ = (10:47:43.32, +11:56:08.4). It has a mean
magnitude of roughly F814W $\approx$ 24.4 and shows variability at the
level of 0.35 mag in the individual ACS images, significantly larger
than the single-image relative magnitude uncertainty of 0.05 mag at that
magnitude. Because of the sparse cadence of our observations, secure
photometry is difficult, but calculated from our two most concurrent
F606W and F814W images (separated by 4.5 days), the object has a 
color of roughly F606W$-$F814=0.7, which would put it near the Cepheid
instability strip in the F606W/F814W color-magnitude diagram \citep[see,
\eg][]{mccommas09}. While our data lack the proper cadence for accurate
phasing, if the object is a Cepheid in BST1047, then with an absolute
magnitude of $M_{\rm F814}=-5.8$ and using the F814W period-luminosity
of \citet{riess19}, the object would have a period of 26 days, roughly
twice the time span of our imaging data, and consistent with the time
variability we see in the source. Without proper imaging cadence, it is
difficult to place strong constraints on the properties of the object,
but if it is a Cepheid, that would also be consistent with the other
signatures of young massive stars that we observe in BST1047.

\begin{figure}[]
\centerline{\includegraphics[width=3.5truein]{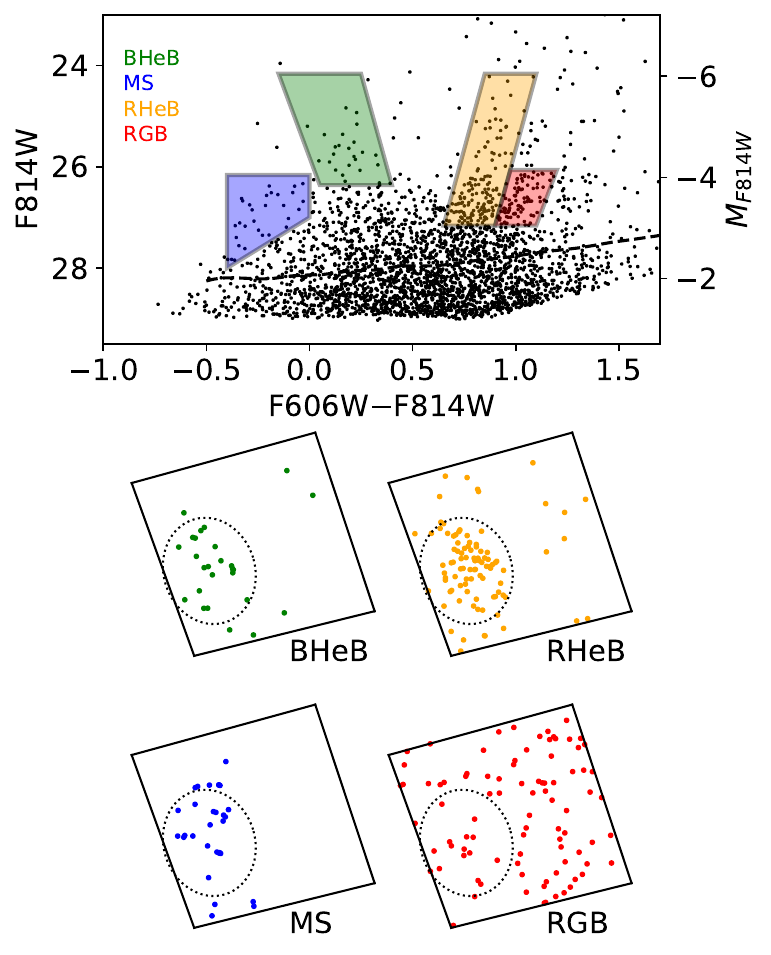}}
\caption{Spatial distribution of point sources in our ACS field, selected by
their position on the color-magnitude diagram. The top panel shows the
CMD for the full ACS field, with regions color coded by their presumed
evolutionary stages. The bottom panels show the ACS field of view, mapping
the point sources corresponding to each selected CMD region. Field orientation
and scale is the same as in Figure~\ref{ACSimage}, and the dotted oval
shows the region containing BST1047.
}
\label{spatplot}
\end{figure}

So far in our analysis we have selected sources spatially, by region
(sources within BST1047 vs those in the surrounding background region)
and considered the CMDs of the regions separately. A complementary
approach is to create a CMD for the entire ACS field, sub-select sources
by their location within this full-field CMD, and ask where these
subsamples are located spatially in the field. We show such an analysis
in Figure~\ref{spatplot}. The panel at the top of the figure shows the
CMD for the whole ACS image, where we have defined regions in the CMD
that highlight the different putative stellar populations discussed so
far. In particular, we highlight the blue and red helium burning
sequences (`BHeB' and `RHeB', respectively), the blue main sequence
region (`MS'), and the location of the older red giant branch population
(`RGB'). Selecting stars that fall in these regions in the CMD, we then
plot in the bottom panels the spatial location of these CMD-selected
sources.

Figure~\ref{spatplot} clearly demonstrates that sources selected from
CMD regions corresponding to young populations --- the MS, BHeB, and
RHeB regions --- are preferentially found within BST1047. In contrast,
objects drawn from the older RGB region are spread much more evenly
across the ACS field, with no preferential clustering in or near
BST1047. These spatial population patterns are consistent with a
scenario in which BST1047 is dominated by recent star formation, with
little or no evidence for an older stellar population. The smoothly
distributed old RGB stars in the field are much more likely to come from
M96's stellar halo, with perhaps some additional contribution from Leo I
intragroup stars \citep[although any such contribution is likely to be
small;][]{watkins14, ragusa22}. Additionally, there is a hint of a weak
gradient in the spatial distribution of RGB stars across the ACS
field. Comparing RGB counts on the western and eastern halves of the
image, we find 54 RGB stars on the west side and 36 on the eastern side,
roughly a $2\sigma$ difference. With M96 located 15\arcmin\ southwest
of the field, this gradient could be tracing the radial dropoff in M96's
halo population, or just a signature of patchiness in M96's halo and/or
the intragroup starlight in the region. For now we leave additional
analysis and a more detailed discussion of the properties of the M96
halo population to a future paper.

\section{The Origin of BST1047+1156}

The detection of massive young stars in the blue and red helium burning
sequences confirms a recent burst of star formation in BST1047, as
originally inferred from the very blue broad-band colors of the galaxy's
integrated light \citep{mihos18bst}. The most luminous stars in these
sequences have absolute magnitudes of $M_{\rm F814W} \approx -5$,
consistent with massive stars ($M_* \approx 6-7$ \Msun) with lifetimes
$<$ 100 Myr, but the sequences continue down to fainter magnitudes
($M_{\rm F814W} \approx -3$) arguing for an extended phase of star
formation extending to at least 300 Myr ago. However, this burst must
have been relatively short lived; the lack of detected \ha\ emission in
BST1047 \citep{donahue95} sets an upper limit on the present-day star
formation rate of $\lesssim 5\times10^{-5}$ \Msun yr$^{-1}$
\citep{mihos18bst}. While we defer a full population modeling analysis
to a future paper, we note here that not only are the red and blue
population sequences consistent with a recent short burst of moderately
metal-poor star formation, such a scenario is also quantitatively
consistent with both the star counts seen in the HST imaging and the
integrated light properties reported in \citet{mihos18bst}. For example,
using the PARSEC stellar population modeling tools \citep[][available at
http://stev.oapd.inaf.it/cmd]{bressan12,marigo17}, a short (50 Myr)
Gaussian burst of star formation of age $\sim 150$ Myr, mass $M_* =
2\times10^{10}$ \Msun, [M/H]=-1.5 and a \citet{kroupa01} initial mass
function yields $\approx$ 40 bright stars in the range $-5 \leq
M_{F814W} \leq -4$ and $\approx$ 180 stars in the fainter range $-4 \leq
M_{F814W} \leq -3$. Performing a similar census of stars in the observed
CMD and correcting for background as shown in \ref{rawcmds} yields 40
$\pm$ 7 stars and 160 $\pm 15$ stars in the brighter and fainter
magnitude range, respectively. This model also yields a total integrated
B magnitude of $M_B=-10.1$ and color of $B-V=0.08$, compared to the
measured values of $M_B=-10.2 \pm 0 .14$ and $B-V=0.14 \pm 0.09$
\citep{mihos18bst}. Thus, all extant data are consistent with a recent,
fading post starburst population in BST1047 that formed within the past
few hundred million years.

Aside from the presence of high mass stars in BST1047, the other notable
feature of its CMD is the lack of any prominent red giant branch
population. While we cannot rule out a modest number of intermediate age
($<$ 5 Gyr) RGB stars, they would need to be very metal poor, with
metallicities [M/H]$\sim -2$, to be hidden within the younger RHeB
sequence. This lack of an old stellar population in BST1047 is in marked
contrast to stellar populations of other types of diffuse star forming
galaxies. The population of dwarf irregulars in the Local Group, while
showing a wide variation of star formation histories, typically show old
populations indicative of extended star formation histories
\citep[\eg][]{grebel97, weisz14}, including even the extremely faint and
diffuse dwarfs such as Leo~T \citep{weisz12} or Leo~P \citep{mcquinn15}.
Another natural comparison would be to the population of blue,
star-forming field low surface brightness galaxies. Resolved stellar
population work in field LSBs has identified the helium burning
sequences from evolving young stars \citep[\eg][]{schombert14,
schombert15} as well as red giant branch stars that trace the older
stellar populations \citep{schombert21}. The lack of RGB stars in
BST1047 thus stands in contrast to the resolved populations in field
LSBs, and more generally the integrated colors of field LSB galaxies are
typically much redder than those of BST1047. For example, the $B-V$
color of BST1047 is 0.14 $\pm$ 0.09 \citep{mihos18bst}, compared to
colors of $B-V \approx$ 0.3 -- 0.6 for field LSBs \citep{mcgaugh94}.
Recent studies show that reddening from dust in LSBs is typically quite
low \citep{junais23}, arguing that the redder colors of these field LSB
galaxies indicate a substantial contribution of light from old stars
relative to what we observe with BST1047. The lack of old stars in
BST1047 likely then rules out scenarios where the galaxy is simply a
extremely low surface brightness outlier in the population of
star-forming field LSB galaxies.

The metallicities of the young stellar populations in BST1047 also argue
against a model in which the object is a pre-existing low mass dwarf
galaxy. Given the low inferred stellar mass for BST1047 (2--4 $\times
10^5$\Msun, this work and \citealt{mihos18bst}), placing it on the
mass-metallicity relationship for dwarf galaxies \citep[\eg][]{kirby13}
would predict a metallicity of [Fe/H]$=-2$, appreciably lower than the
metallicity inferred from the analysis shown in
Figure~\ref{youngtracks}.

A more likely scenario for BST1047 is that it formed during a tidal
interaction between galaxies within the Leo I Group. Tidal interactions
can strip gas from the gas-rich outer disks of spiral galaxies,
expelling that gas into the surrounding environment. Concurrently with
the stripping the gas in the tidal debris can be collisionally
compressed, leading to a burst of star formation and, potentially, to
the formation of a tidal dwarf galaxy \citep[\eg][]{duc00, bournaud06,
lelli15}. In this aspect, BST1047 may be most similar to (albeit fainter
and much more diffuse than) tidal dwarf candidates found in the M81
group \citep{durrell04, mouhcine09, chiboucas13}, which also appear to
lack old stellar populations. Because tidal dwarfs form from
pre-enriched material stripped from a larger host galaxy, these objects
should also be elevated in metallicity compared to regular dwarf
galaxies of the same mass \citep{duc98, weilbacher03}, just as we find
for BST1047. Indeed, the metallicity of the young stars in BST1047 is
comparable to that found in the outskirts of large spirals
\citep[\eg][]{zaritsky94, vanzee98, berg20}, and is also distinct from
the higher, solar-like metallicities found in the Leo Ring to the north
\citep{corbelli21}. Thus, the most likely origin for BST1047 is from gas
that was stripped from the outer disk of the spiral galaxy M96, as also
suggested by the distorted tidal HI morphology of M96's outer disk
\citep[Figure~\ref{optHI} and][]{oosterloo10}.

Such a scenario would also explain the presence of massive young stars
forming in an object with such low gas density, well below that more
typically found in star forming environments \citep[\eg][]{bigiel08,
bigiel10, wyder09}. If BST1047 formed during a tidal encounter, the
initial compression of gas in the tidal caustics would have led to much
higher gas densities capable of driving a weak starburst now traced by
the young populations in BST1047. Subsequent tidal or ram-pressure
stripping of the object, coupled perhaps with energy input from stellar
winds and supernovae from the evolving starburst population, could have
then dissociated the molecular gas and left BST1047 with a very diffuse
ISM. The peak column density in BST1047 today is very low \citep[$\sim$
1 \Msun\ pc$^{-2}$][]{mihos18bst}, and likely incapable of fueling any
additional star formation. While the relatively large beam size of the
HI data leaves open the possibility of pockets of high density gas on
small scales, recent CO observations of BST1047 have failed to detect
molecular gas in the system \citep{corbelli23}, although those studies
did not survey BST1047's full HI extent. Nonetheless, there is no
evidence for any current, ongoing star formation in BST1047 today.

The ultimate fate of BST1047 remains unclear. Its HI morphology
(Figure~\ref{optHI}) shows streamers of HI extending to the southeast
towards M96, likely a signature of tidal stripping of BST1047, or
perhaps ram-pressure stripping from hot gas in M96's halo or the group
environment. BST1047's very low density makes it susceptible to
stripping processes, particularly if it is a tidal dwarf with no
cocooning halo of dark matter to keep it bound. \citet{mihos18bst} used
the observed HI kinematics to show that BST1047's dynamical mass and
baryonic mass were comparable ($\sim 5-6 \times 10^7$ \Msun), providing
support for the idea that the object lacks dark matter, as expected for
tidal dwarfs. If BST1047 is in the process of being disrupted by the
environment of the Leo I Group, we may be catching this object in a very
transitory phase, with its fading post-starburst population soon to be
stripped and expelled into the group environment. As such, the galaxy
may be a prime example of a ``failing'' tidal dwarf, born in the tidal
debris of a recent encounter, but lacking sufficient mass to overcome
the destructive dynamical processes found within the group environment.

\section{Summary}

We have used deep {\sl Hubble} ACS imaging in F606W/F814W to study the
resolved stellar populations in the gas-rich ultradiffuse object
BST1047+1156 in the Leo I Group. At zero color, our photometry reaches
limiting magnitudes of F606W$_{\rm lim}$=28.7 and F814W$_{\rm lim}$=28.2,
extending two magnitudes down the red giant branch at the 11.0 Mpc
distance of the Leo I Group. We clearly detect the stellar population
associated with BST1047, identifying the red and blue helium burning
sequences expected from an evolving population of massive stars. We also
find an excess of fainter blue stars likely to be slightly less massive
stars still on the main sequence. The distribution of color and
luminosity of stars in BST1047 are consistent with a modestly metal poor
stellar population ([M/H] $\sim -1.0$ to $-1.5$) with ages of a few
hundred million years, consistent with the integrated colors and surface
brightness measured in ground based imaging \citep{mihos18bst}.

However, we find no trace of a red giant branch sequence in the stellar
populations of BST1047 despite going sufficiently deep to detect such
stars. This lack of an old stellar population argues strongly against
scenarios in which BST1047 is a long lived LSB galaxy that has merely
had a weak burst of star formation due to interactions
within the group environment. Instead,
the combination of its exclusively young and moderately metal-poor
stellar populations, its diffuse nature, and its disturbed HI morphology
argues that we are seeing a transient object, likely formed from gas
recently stripped from the outer disk of M96 due to tidal forces at work
within the group environment. These tidal forces continue to strip gas
and stars away from BST1047 today, feeding the the intragroup stellar
population of the Leo I Group. BST1047 is thus likely to be a failing
tidal dwarf, formed from the tidal debris of M96 but with such low
density that it is destined to ultimately disperse into the intragroup
population of the Leo I group.

Finally, in the environment surrounding BST1047, we also clearly detect
red giant stars in the stellar halo of M96. These stars are distributed
fairly uniformly across the ACS field of view, showing no spatial
correlation with the location of BST1047. From the location of the red
giant sequence on the color-magnitude diagram we infer a moderately low
stellar metallicity of [M/H] $\approx -1.3 \pm 0.2$. These data probe
the stellar populations in the galaxy's outer halo at the extremely
large projected radial distance of 50 kpc, and we plan a future paper
incorporating the data from the adjacent WFC3 parallel field to study the
properties of M96's outer stellar halo in more detail.

\begin{acknowledgments}

The authors would like to thank Christian Soto and Norman Grogin for
their help with planning and refining the HST observations, as well as
the anonymous referee for helpful suggestions that improved the
presentation of our results. This research is based on observations made
with the NASA/ESA Hubble Space Telescope for program \#GO-15258 and
obtained at the Space Telescope Science Institute (STScI). STScI is
operated by the Association of Universities for Research in Astronomy,
Inc., under NASA contract NAS5-26555. Support for this program was
provided by NASA through grants to J.C.M. and P.R.D. from STScI. A.E.W.
acknowledges support from the STFC through grants ST/S00615X/1 and
ST/X001318/1.

\end{acknowledgments}

\facility{HST (ACS)}. The {\sl Hubble Space Telescope} imaging data used
in this study can be accessed at the Mikulski Archive for Space
Telescopes (MAST) at the Space Telescope Science Institute via
\dataset[DOI: 10.17909/300a-ns91]{https://doi.org/10.17909/300a-ns91}.

\software{
astropy  \citep{astropy13, astropy18, astropy22}, 
DOLPHOT \citep{dolphin00},
numpy \citep{numpy},
matplotlib \citep{matplotlib},
scipy \citep{scipy},
}

\bibliographystyle{aasjournal}

\end{document}